# Giant second-order nonlinearity in twisted bilayer graphene

Junxi Duan[1,#,*], Yu Jian[1,#], Yang Gao[2*], Huimin Peng[1], Jinrui Zhong[1], Qi Feng[1], Jinhai Mao[3], Yugui Yao[1*]

[1]Key Laboratory of Advanced Optoelectronic Quantum Architecture and Measurement (MOE), School of Physics, Beijing Institute of Technology, Beijing, China

[2]Department of Physics, University of Science and Technology of China, Hefei, Anhui, China

[3]School of Physical Sciences and CAS Center for Excellence in Topological Quantum Computation, University of Chinese Academy of Sciences, Beijing, China

[#]These authors contributed equally: J. Duan, Y. Jian

[*]Corresponding authors: junxi.duan@bit.edu.cn, ygao87@ustc.edu.cn, ygyao@bit.edu.cn

In the second-order response regime, the Hall voltage can be nonzero without breaking the time-reversal symmetry, as long as the system is noncentrosymmetric. There are multiple mechanisms with different scaling rules that contribute to the nonlinear Hall effect (NLHE). The intrinsic contribution is closely related to the Berry curvature dipole and has been extensively investigated recently. The study of the extrinsic contribution, however, is scarce, although it can enter the NLHE even in the leading order. Here, we report a giant nonlinear transport response in TBG, in which the intrinsic mechanism is forbidden. The magnitude and direction of the second-order nonlinearity can be effectively tuned by the gate voltage. The peak value of the second-order Hall conductivity close to the full filling of the moiré band reaches $8.76\ \mu mSV^{-1}$, four-order larger than those detected in $WTe_2$. The observed giant second-order nonlinearity can be understood from the collaboration of the asymmetric scattering of electrons off the static (Coulomb impurities) and dynamic disorders (phonons) in noncentrosymmetric crystals. It is mainly determined by the skew-scattering contribution from impurities at 1.7 K. The skew-scattering from phonons has a much larger coupling coefficient as suggested by the scaling results, and becomes as important as the impurity contribution as the temperature rises.

**Our observations demonstrate the potential of TBG in studying nonlinear response and possible rectification applications.**

Since the discovery of the Hall effect by Edwin H. Hall in 1879, the family of the Hall effects has been largely expanded. The deep understanding of the quantum nature of the transverse velocity of electron enriches the condensed matter physics with the concepts of Berry curvature and topological Chern number[1-3]. The Hall effects provide direct ways to the detection of the topology of the electronic band structures, such as the quantum Hall effect and quantum anomalous Hall effect[4-6]. To get nonzero Hall signal in the linear response regime, the time-reversal symmetry should be broken by external magnetic field or in magnets. Recently, a second-order Hall response named nonlinear Hall effect (NLHE) has attracted intense interest[7-17]. Distinct from the linear Hall effect, the NLHE does not require time-reversal symmetry breaking but inversion symmetry breaking. The first observation of the NLHE in $WTe_2$ demonstrated its close relation to the high-order topological properties, the Berry curvature dipole (BCD)[7,10,11]. Determined by the topology of the electronic band structure, the BCD contribution is disorder-independent. However, disorder always plays an indispensable role in transport response. Theories and experiments have already shown that the extrinsic (disorder-induced) contribution is comparable to the intrinsic (disorder-free) one in the linear Hall effect[2,18,19]. In the NLHE, disorder is more important since it can enter the NLHE even in the leading order[20-22]. Although the disorder contributions to the NLHE have been intensively discussed in theory, experimental investigations are scarce.

To study the disorder contributions, it is essential to segregate them from the intrinsic one. Consequently, systems with certain point group symmetry are required, since the intrinsic contributions are fully determined by the BCD[7]. From symmetry analysis, two-dimensional systems with point groups such as $C_3$, $C_{3h}$, and $C_{3v}$ forbid intrinsic contributions to the NLHE[22]. Among the potential candidates, twisted bilayer graphene (TBG) has attracted intense interest. A series of exotic properties have been reported in this system in linear-response regime, including correlation-induced insulator and superconductivity at certain magic angles where Coulomb interactions dominate over the kinetic energy of electrons[23-26], and topological Chern insulator, ferromagnetism, and quantum anomalous Hall effect stemming from the spin- and valley-resolved moiré bands[27-30]. Although pristine TBG has a $D_6$ symmetry, it is reduced to $C_3$ by the hexagonal boron nitride (hBN) substrate. We identify TBG as an ideal system for the study of disorder contributions to NLHE with the following advantages. First, in TBG, a prominent characteristic is the formation of moiré superlattice which folds the Brillouin zone, creating moiré superlattice gaps and modulating the velocity of the Dirac cone[31,32]. These moiré superlattice gaps are crucial to the NLHE which is generally stronger near the band edge. Second, the chemical potential in TBG is gate-tunable, and all the NLHE contributions are chemical-potential sensitive. Last, the disorder contributions to NLHE favor high mobility which is a typical feature of graphene systems. Here, we report a giant nonlinear transport response in TBG. An extremely large second-order Hall conductivity is observed close to the moiré superlattice gaps, orders higher than the values obtained in $WTe_2$ and twisted $WSe_2$ systems[10-12]. The scaling law of the NLHE extracted from temperature-dependent measurements well matches the multivariable skew-scattering contributions scheme. Surprisingly, we find that the electron-phonon scattering has a much larger coupling coefficient than that of the electron-impurity scattering, and as the temperature rises, it can gradually reach the same order as the latter.

TBG samples in this study are prepared following the "tear-and-stack" method (See Methods)[33-35]. The twisted angle is well controlled to be near the magic angle during the stacking process. Figure 1a presents a sketch of the device structure and measurement setup with the defined $x$ and $y$ directions. Figure 1b and 1c show the gate dependence of the longitudinal resistance measured at 300 K and 1.7 K, respectively. The two prominent peaks observed at 1.7 K identify the positions of the moiré superlattice gaps. From the carrier density $n_s$ required to fulfill the moiré flat band, the actual twisted angle is estimated to be about $1.22°$. As conventional, we will use the moiré filling factor $v = 4n/n_s$, corresponding to the number of electrons per moiré unit, to denote the position of the chemical potential. The existence of insulating states near the half filling of the moiré band, $v = \pm 2$, at 1.7 K as well as the broad peak observed at 300 K indicate the formation of the moiré flat bands[24]. At the lowest temperature of our measurements (1.7 K), however, no superconducting behavior is observed in current device.

To measure the nonlinear response, an AC current with fixed frequency (17.777Hz) is injected into the Hall bar device (See Supplementary). The longitudinal and transverse voltage drops at both the base and second-harmonic frequencies are detected simultaneously with standard lock-in amplifiers (See Methods). The relations between the second-harmonic transverse voltage $V_y^{2\omega}$ and the injected current $I_x$ under different gate voltages are measured at 1.7 K, with no external magnetic field applied (see Supplementary). All the $V_y^{2\omega}$ depend nonlinearly on the injected current $I_x$. Since the resistance of the sample is tuned by the gate voltage as well, it is convenient and clear to plot the $V_y^{2\omega}$ against the square of the longitudinal voltage drop $V_x^2$. As plotted in Fig. 2a, it is obvious that all the second-harmonic transverse voltage $V_y^{2\omega}$ linearly depend on $V_x^2$. Moreover, the direction of $V_y^{2\omega}$ does not depend on the direction of the injected current, shown in Fig. 2b, confirming that it is a second-order nonlinear effect. Figure 2c presents the moiré-filling-factor dependence of $V_y^{2\omega}$ measured under $I_x = 1\mu A$. The magnitude of $V_y^{2\omega}$ strongly depends on the moiré filling factor. In addition, the sign of $V_y^{2\omega}$ varies as $\nu$ is changed. It switches several times when $\nu$ is tuned from -5 to +5. The moiré-filling-factor dependence of $V_y^{2\omega}$ with the current direction rotated by $90°$ is also measured and presented in Fig. 2c. Obviously, the $V_y^{2\omega}$ after rotation shows similar zero-crossings. Its magnitude is slightly smaller than the one before rotation, but the direction is opposite. In general, a second-order nonlinear effect can be written as $j_\alpha^{2\omega} = \chi_{\alpha\beta\beta} E_\beta E_\beta$, where $j_\alpha^{2\omega}$ is the second-order current in direction $\alpha$ driven by the electric field $E_\beta$ in the direction $\beta$, $\chi_{\alpha\beta\beta}$ is the rank-three tensor for the second-order conductivity[7,21]. Figure 2d presents the moiré-filling-factor dependence of the second-order Hall conductivity $\chi_{yxx}$ measured at 1.7 K. $\chi_{yxx}$ shows two striking peaks below and above the moiré flat bands, close to the moiré superlattice gaps. The peak value of $\chi_{yxx}$ reaches an extremely large value, 8.76 μmSV$^{-1}$, orders higher than the values reported in WTe$_2$ and twisted WSe$_2$ systems[10-12].

We next explore the microscopic mechanisms of the observed nonlinear effect. In general, $\chi_{yxx}$ consists of both the intrinsic and the extrinsic contributions[21,22]. While the former is well determined by the Berry curvature dipole and the transport relaxation time[7,11], the latter is more complicated and relies on not only the type of the scattering sources but also its strength and shape[20-22]. Phenomenologically, the intrinsic contribution can only contribute to the transverse transport, while the extrinsic one can lead to the longitudinal transport as well[20,22]. For TBG on hBN substrate, the $C_3$ point group symmetry forbids any intrinsic contribution. Therefore, the second-order response in TBG always have the extrinsic origin, and hence the longitudinal and transverse nonlinear transport should both exist and have the similar order of magnitude in general. Our data confirm such a property and hence the extrinsic origin (See Supplementary). We will focus on the transverse response below.

According to theory[21], the general scaling law of the rank-three tensor $\chi_{\alpha\beta\beta}$ can be written as $\chi_{\alpha\beta\beta} = A_1\sigma^3 + A_2\sigma^2 + A_3\sigma$, where $A_i$ ($i = 1, 2, 3$) are the scaling parameters, $\sigma$ the conductivity. Since we measure the transverse voltage drop, the scaling law is rewritten as

$$\frac{E_y^{2\omega}}{E_x^2} = \frac{\chi_{yxx}}{\sigma} = A_1\sigma^2 + A_2\sigma^1 + A_3\sigma^0 \tag{1}$$

where $E_y^{2\omega} = V_y^{2\omega}/W$ and $E_x = V_x/L$ can be deduced from the measured voltage drops, $W$ and $L$ the width and length of the channel, respectively. We varied the conductivity by changing the device temperature, with the gate voltage fixed. Figure 3 presents the typical results measured at $V_g = -0.8$ V, corresponding to $\nu = -0.42$. Figure 3a shows the $V_x^2 - V_y^{2\omega}$ relation at different temperatures ranging from 1.7 K to 80 K. It is obvious that $V_y^{2\omega}$ depends linearly on $V_x^2$ at all temperatures. The slope of the linear dependence, $\frac{E_y^{2\omega}}{E_x^2} = \frac{L^2}{W}\frac{V_y^{2\omega}}{V_x^2}$, decreases with increasing temperature (Fig. 3b). On the other hand, the two-dimensional conductivity $\sigma$ has quite similar temperature dependence (Fig. 3c). We then plot $E_y^{2\omega}/E_x^2$ against $\sigma$ and find perfect parabolic dependence (Fig. 3d). Referring to equation (1), the parabolic dependence of $E_y^{2\omega}/E_x^2$ on $\sigma$ indicates that the measured nonlinear Hall effect has all the three components corresponding to the three terms on the right of the equation. This is totally different from previous observations in WTe$_2$, and twisted WSe$_2$, where only the first and third terms are identified[10-12].

The scaling results shed light on the mechanisms that contribute to the observed NLHE. As introduced by ref.[21], the extrinsic contributions to the second-order conductivity $\chi_{yxx}$ can be classified into three different types, i.e., side-jump, Gaussian skew-scattering, and non-Gaussian skew-scattering. For each scattering mechanism, there are further two different scattering sources: static disorders (e.g., Coulomb impurities) and dynamic disorders (e.g., phonons). Each of the three scaling parameters $A_i$ involves a mixture of several different mechanisms and scattering sources[21]. Since the scattering with phonon is sensitive to temperature, it can be used to tune the conductivity and hence extract the $A_i$. To identify the dominant mechanisms, examining the moiré-filling-factor dependence of the scaling parameters is helpful. Figure 4a and 4b plot two typical $E_y^{2\omega}/E_x^2$ against $\sigma$ relations measured at $\nu = 0.78$ and $\nu = -1.31$, respectively. Obviously, the scaling law at different moiré filling factors can be well captured by equation (1). The three components in equation (1) depend differently on the moiré filling factor. We then extract the scaling parameters at different gate voltages. It should be mentioned that the scaling is not possible around the half-filling of the moiré band, $\nu = \pm 2$. The reason may be the strong electron-electron correlations, which cause a metal-insulator transition at low temperature. Its effect on the nonlinear response is not included in equation (1) which is deduced from a single-particle point of view. However, this will do no harm to our discussion since the strongest response is observed around the moiré superlattice gaps away from the half fillings.

To explore the coupling coefficients of different types of contributions, we should take the zero-temperature conductivity into account in the obtained scaling parameters, to discount the effect of the disorder strength[21]. For this purpose, we identify the conductivity at 1.7 K as the zero-temperature conductivity $\sigma_0$ since the corresponding energy scale $\sim 14$ μeV is already much smaller than the phonon energy scale. Figure 4c plots the modified scaling parameters, $A_1\sigma_0^2$, $A_2\sigma_0$, and $A_3$ against the moiré filling factor. The result show three interesting features that directly illuminate the origin of the giant NLHE. First, the three scaling parameters have the same orders of magnitude. Since the non-Gaussian skew-scattering only affects $A_1\sigma_0^2$, such an approximately equal magnitude of scaling parameters then suggests that the non-Gaussian skew-scattering should only play a secondary role. In this case, $A_1\sigma_0^2$, $A_2\sigma_0$, and $A_3$ do not depend on the concentration of the scatterings any more, but only rely on the source and shape of the scattering events. Secondly, all three scaling parameters have quite similar moiré-filling-factor dependence. Most strikingly, their peak structures are well synchronized: they all contain two extremely large peaks, one above and one below the moiré superlattice gaps. This feature further excludes the side-jump mechanism as the dominant contribution near the large peaks, since $A_1\sigma_0^2$ is not affected by the side jump at all and would develop a peak at different positions if the side-jump contribution were the dominant. Finally, near both peaks, we find that $A_1\sigma_0^2:A_2\sigma_0:A_3 \approx 1:-2:1$. Referring to ref.[21], this special ratio strongly hints that the phonons dominate the three scaling parameters, as the impurities favor stronger $A_1\sigma_0^2$ but weaker $A_2\sigma_0$ and $A_3$.

To compare the skew-scattering distributions from different scattering sources, we can rewrite equation (1) in the following form: $\chi_{yxx} = \sigma\left[C_i\left(\frac{\sigma}{\sigma_0}\right)^2 + C_{i-ph}\left(\frac{\sigma}{\sigma_0}\right)\left(1-\frac{\sigma}{\sigma_0}\right) + C_{ph}\left(1-\frac{\sigma}{\sigma_0}\right)^2\right]$, ignoring the side-jump contributions, where the three coupling coefficients are independent on temperature, $C_i$ is from the electron-impurity scattering, $C_{i-ph}$ from the mixed scattering off both phonons and impurities, and $C_{ph}$ from the electron-phonon scattering. As shown by our data, $\sigma$ and $\sigma_0$ are of the same order, so the additional factor $\sigma$ suggests that a large mobility can lead to the giant second-order conductivity. At zero temperature, we have $\sigma = \sigma_0$. While the phonons vanish, the second-order nonlinearity is solely from impurities, i.e. the skew scattering from impurities at the third- (non-Gaussian) and forth-order (Gaussian) in perturbation theory[21]. At finite temperatures, phonons emerge and enter the scattering events. Unlike impurities, phonons only enter the skew-scattering at the forth-order in perturbation theory. According to the scaling results, the coupling coefficient from the electron-phonon scattering is much larger the one from the electron-impurity scattering. Therefore, as the temperature increases, the electron-phonon contribution will also increase and gradually compete with the electron-impurity contribution (Fig. 4d).

At last, we would like to comment on the intrinsic contribution. Ideally, TBG on substrate holds the C₃ point symmetry. However, strain is quite common in TBG caused by the "tear-and-stack" processes. With uniaxial strain, the point symmetry of the system is further reduced to C₁, allowing nonzero intrinsic contribution. Recent theoretical studies have evaluated the intrinsic contribution to NLHE in a strained TBG[14,36]. Under 0.1% uniaxial strain, the peak value of the coefficient of the intrinsic contribution can only reach ~1 μmV⁻¹, three orders smaller than the measured extrinsic coefficients, that its contribution to NLHE can always be neglected.

In conclusion, we observed giant nonlinear Hall response in twisted bilayer graphene under zero magnetic field. The magnitude and direction of the second-order conductivity can be effectively controlled by gating. The second-order Hall conductivity around the moiré superlattice gaps reaches 8.76 $\mu mSV^{-1}$, orders larger than those reported in WTe$_2$ and twisted WSe$_2$. The scaling law deduced from temperature-dependent measurements consists of multiple components, well matches the multivariable skew-scattering contributions scheme. In addition, the electron-phonon scattering has a much larger coupling coefficient than that from the electron-impurity scattering. As the temperature rises, the electron-phonon contribution also increases and becomes competitive with the electron-impurity contribution. Our results demonstrate the potential of TBG in deeper understanding of the extrinsic mechanisms in the NLHE.

NOTE: During the preparation of this manuscript, we became aware of a recent article reporting of a second-order Hall conductivity with its magnitude similar to ours in graphene encapsulated by hexagonal boron nitride devices[37].

**Method:**

**Device fabrication**

Twisted bilayer graphene devices were prepared by tear-and-stack method. Single layer graphene and hexagonal boron nitride (hBN) flakes were mechanically exfoliated from bulk crystals (hBN from HQ graphene) onto silicon substrates. The hBN flake was first picked up by a stamp consisting of polycarbonate (PC) on polydimethylsiloxane (PDMS). Then, it was used to pick up half of a single-layer graphene. The other half of graphene was rotated by a certain angle (1.2 degree for the device discussed in the main text) by high-resolution-motor-drive stage before it was picked up. After that, a bottom hBN flake was picked up. The stack was finally released onto a thin graphite flake which serves as the gate. To make electrical contact, edge contact method was employed. Cr/Pd/Au layers with thickness of 3nm/5nm/60nm were prepared by e-beam evaporation under a pressure about $1 \times 10^{-6}$ Pa. The devices were then patterned into standard Hall bars, followed by reactive ion etching using $CHF_3$ and $O_2$ mixing gas.

**Transport measurements**

The transport measurements were carried out in a Physical Property Measurement System (PPMS) (Dynacool, Quantum Design). The lowest temperature is 1.7 K. An AC current was applied to the device. The longitudinal and transverse voltage drops were simultaneously measured at fundamental and the second-harmonic frequencies with lock-in amplifiers. The two-dimensional conductivity was obtained as $\sigma = \frac{I_x}{V_x}\frac{L}{W}$, where $L$ and $W$ are the length and width of the Hall bar device, respectively. The moiré filling factor was tuned by a gate voltage. To confirm the second-order response, the magnitude of the AC current was swept from zero to a certain current ranging from 1 μA to 5 μA depending on the resistance of the sample under gating. For the temperature-dependent measurements, a gate voltage was first set and kept constant. Then, at each temperature, the second-order response under different AC current excitation was recorded.

To get the second-order conductivity, we should rewrite $j_\alpha^{2\omega} = \chi_{\alpha\beta\beta} E_\beta E_\beta$ since we measure the voltage drop instead of the current. Considering the external electric field oscillating with frequency $\omega$, it leads to

$$E_\alpha^{2\omega} = \frac{\chi_{\alpha\beta\beta}}{\sigma} E_\beta E_\beta = \frac{\chi_{\alpha\beta\beta}}{\sigma} E_{\beta 0} E_{\beta 0} \sin^2\omega t = \frac{\chi_{\alpha\beta\beta}}{2\sigma} E_{\beta 0} E_{\beta 0} [1 - \sin 2\omega t]$$

where $E_{\beta 0}$ denotes the magnitude of the oscillating electric field. In the nonlinear measurement, we record the voltage drop at the frequency $2\omega$. Thus,

$$E_{\alpha 0}^{2\omega} \sin 2\omega t = -\frac{\chi_{\alpha\beta\beta}}{2\sigma} E_{\beta 0} E_{\beta 0} \sin 2\omega t$$

where $E_{\alpha 0}^{2\omega}$ is the magnitude of the measured electric field at $2\omega$. For the transverse response, it is

$$E_{y0}^{2\omega} \sin 2\omega t = -\frac{\chi_{yxx}}{2\sigma} E_{x0}^2 \sin 2\omega t$$

Therefore, the second-order Hall conductivity derived from the lock-in measurements is

$$\chi_{yxx} = -2\sigma \frac{E_{y0}^{2\omega}}{E_{x0}^2} = -2\sigma \frac{L^2}{W} \frac{V_y^{2\omega}}{V_x^2}$$

where $V_y^{2\omega}$ and $V_x^2$ are the magnitudes of the transverse and longitudinal voltage drops at frequency $2\omega$ and $\omega$, respectively, which are the values measured by the lock-in amplifiers, $W$ and $L$ the width and length of the channel, respectively.

**Acknowledgements**

The research was supported by the National Key R&D Program of China (Grant Nos. 2020YFA0308800, 2019YFA0308402), the National Natural Science Foundation of China (Grants Nos. 61804008, 11734003), the Beijing Natural Science Foundation (Z190006), the Strategic Priority Research Program of Chinese Academy of Sciences (Grant No. XDB30000000). The fabrication was supported by Micronano center of Beijing Institute of Technology. J.D. acknowledges the support of the Beijing Institute of Technology Research Fund Program for Young Scholars. Y.G. acknowledges the support of the Fundamental Research Funds for the Central Universities, China (Grant No. WK2340000102). The authors thank Tingxin Li for fruitful discussions.

**Author contributions**

J.D. and Y.Y. conceived and designed the experiments. Y.J. fabricated the devices with the help from H.P., J.Z., and Q.F. J.D. and Y.J. performed the measurements and analyzed the data with the help from H.P., J.Z., and Q.F. Y.G. performed the theoretical studies. J.D., Y.G., J.M., and Y.Y. wrote the manuscript. All the authors discussed the results and commented on the manuscript.

**Competing interests**

The authors declare no competing interests.

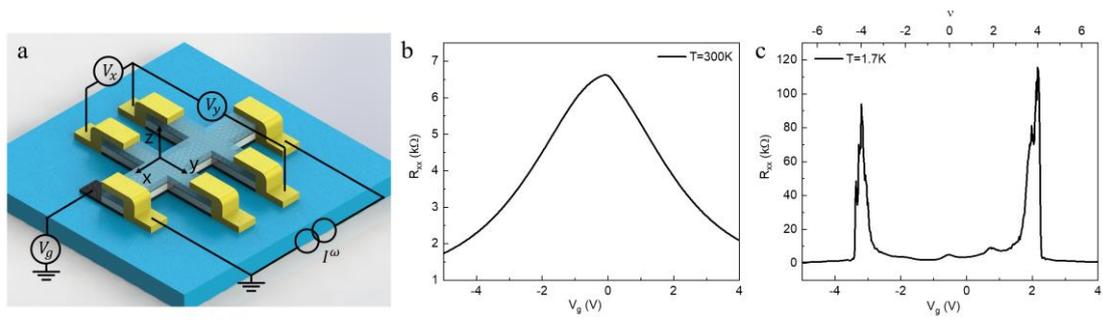

**Fig.1 Basic characterization of TBG device. a**, sketch of the structure of TBG device and the measurement setup. A graphite flake under the bottom hBN is used as the gate. **b,c**, gate dependence of longitudinal resistance measured at 300 K (**b**) and 1.7 K (**c**).

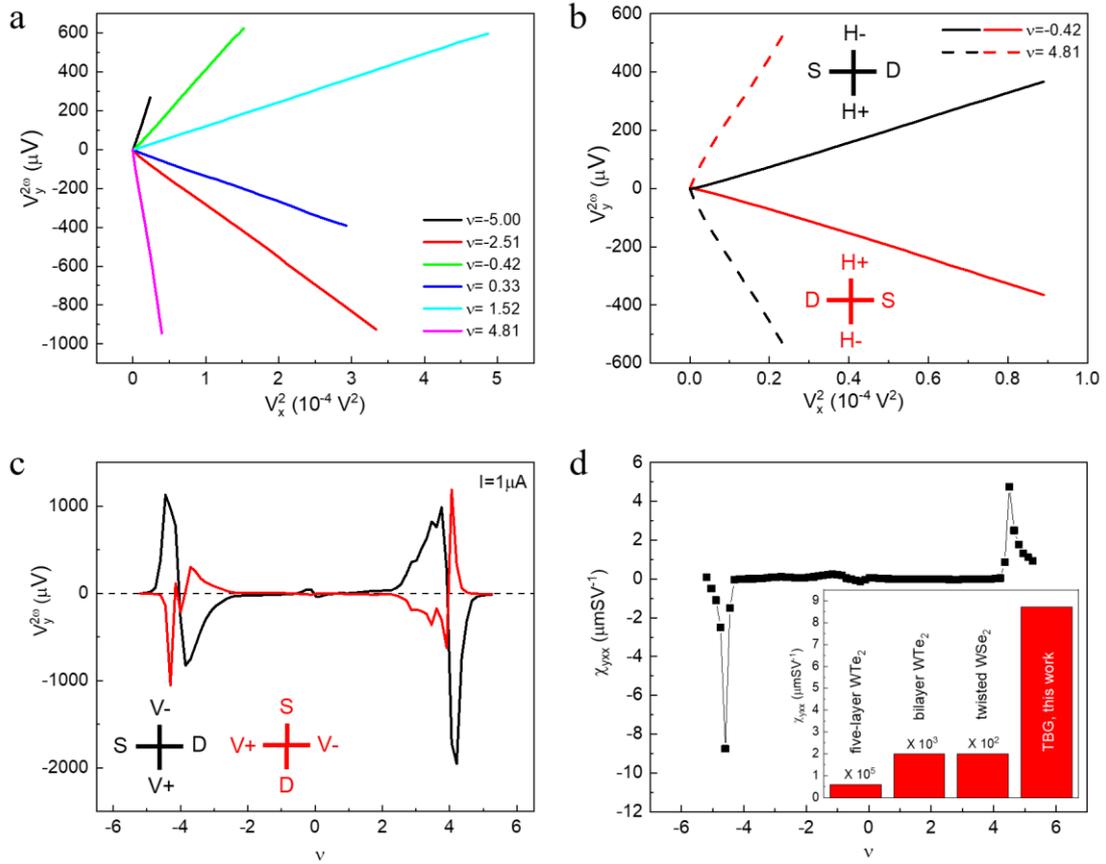

**Fig.2 Nonlinear Hall response. a**, $V_y^{2\omega}$ depends linearly on $V_x^2$ at different moiré filling factors. **b**, $V_y^{2\omega}$ changes sign when the current direction is reversed, measured at $\nu = -0.42$ (solid lines) and $\nu = 4.81$ (dash lines). The insets sketch the measurement geometry with forward (black) and backward (red) current. **c**, moiré-filling-factor dependence of $V_y^{2\omega}$ detected under 1 µA excitation, before (black) and after (red) the 90° rotation of the current direction. The insets sketch the measurement geometry. **d**, moiré-filling-factor dependence of the second-order Hall conductivity. The inset compare the second-order Hall conductivity measured in current work with the values in the existing reports. $\chi_{yxx}$ for  Data in **a-d** were collected at 1.7 K.

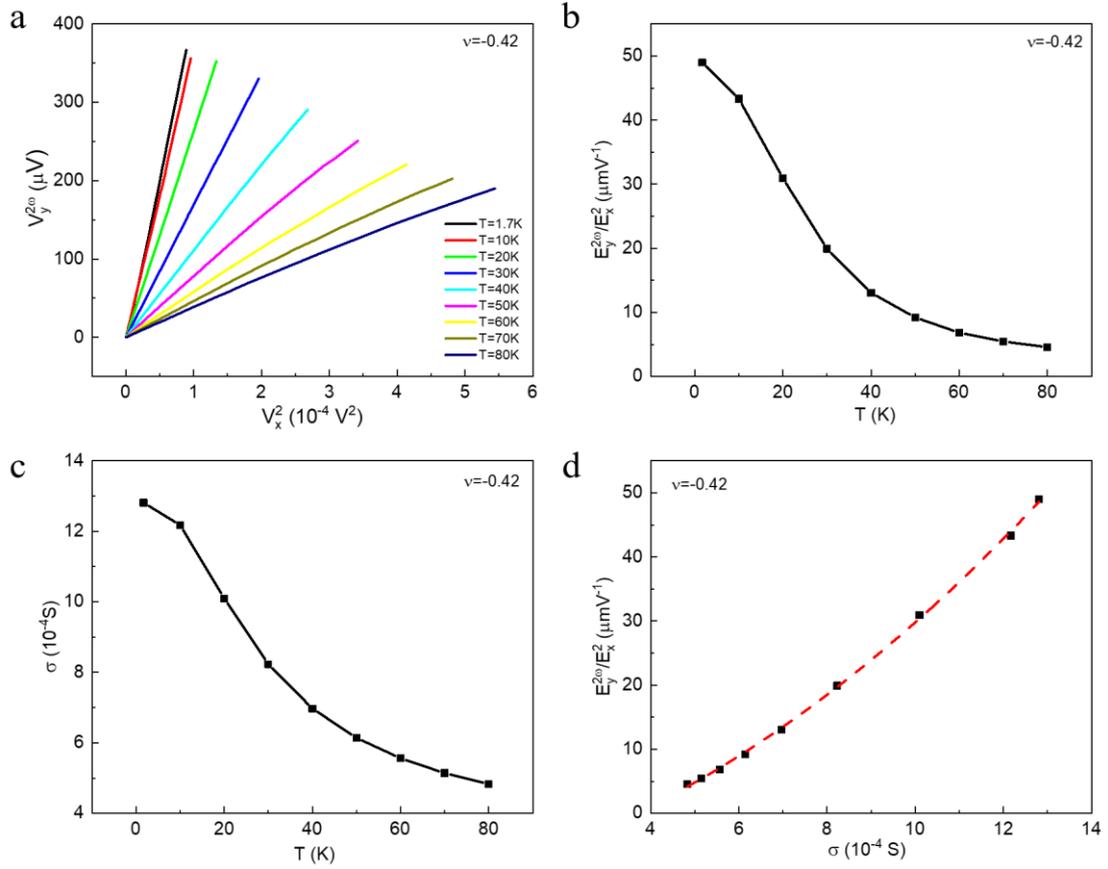

**Fig.3 Temperature dependence of nonlinear Hall effect. a**, $V_y^{2\omega}$ depends linearly on $V_x^2$ at different temperatures ranging from 1.7K to 80 K with $\nu = -0.42$. **b,c**, temperature dependence of $\frac{E_y^{2\omega}}{E_x^2}$ (**b**) and two-dimensional conductivity $\sigma$ (**c**). **d**, $\frac{E_y^{2\omega}}{E_x^2}$ as a function of $\sigma$. The lines **b** and **c** are guides to the eye. The dash line in **d** is a parabolic fit to the experimental data.

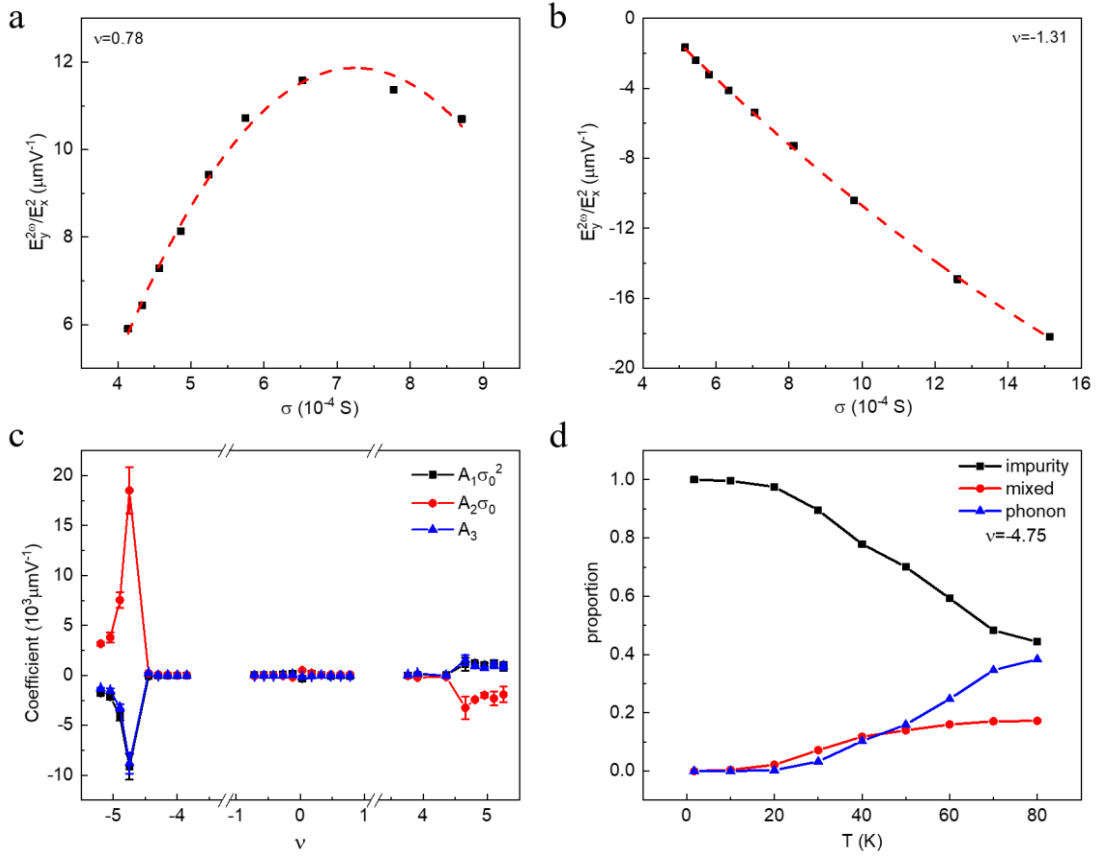

**Fig.4 Gate dependence of scaling parameters. a,b**, $\frac{E_y^{2\omega}}{E_x^2}$ as a function of $\sigma$ obtained at $\nu = 0.78$ (**a**) and $\nu = -1.31$ (**b**). **c**, gate dependence of the modified scaling parameters, $A_1\sigma_0^2$, $A_2\sigma_0$, and $A_3$. **d**, temperature dependence of the proportion of the contribution to second-order conductivity from electron-impurity, mixed, and electron-phonon scatterings deduced from the scaling results measured at $\nu = -4.75$. The dash lines in **a** and **b** are parabolic fit to the experimental data. The solid lines in **c** and **d** are guide lines to the eye.

# Supplementary Materials

**Giant second-order nonlinearity in twisted bilayer graphene**

Junxi Duan[1,#,*], Yu Jian[1,#], Yang Gao[2*], Huimin Peng[1], Jinrui Zhong[1], Qi Feng[1], Jinhai Mao[3], Yugui Yao[1*]

[1]Key Laboratory of Advanced Optoelectronic Quantum Architecture and Measurement (MOE), School of Physics, Beijing Institute of Technology, Beijing, China

[2]Department of Physics, University of Science and Technology of China, Hefei, Anhui, China

[3]School of Physical Sciences and CAS Center for Excellence in Topological Quantum Computation, University of Chinese Academy of Sciences, Beijing, China

[#]These authors contributed equally: J. Duan, Y. Jian

[*]Corresponding authors: junxi.duan@bit.edu.cn, ygao87@ustc.edu.cn, ygyao@bit.edu.cn

1. Nonlinear response under different frequencies

In the main text, nonlinear response was driven by AC current with frequency 17.777 Hz. We have also measured the response under different driving frequencies. Figure S1 shows the second-harmonic Hall voltage $V_y^{2\omega}$ as a function of $V_x^2$ under different driving frequencies. No obvious frequency dependence is observed in the frequency range we applied (13.777 – 33.777 Hz). This is incompatible with capacitive coupling effect which strongly depends on driving frequency.

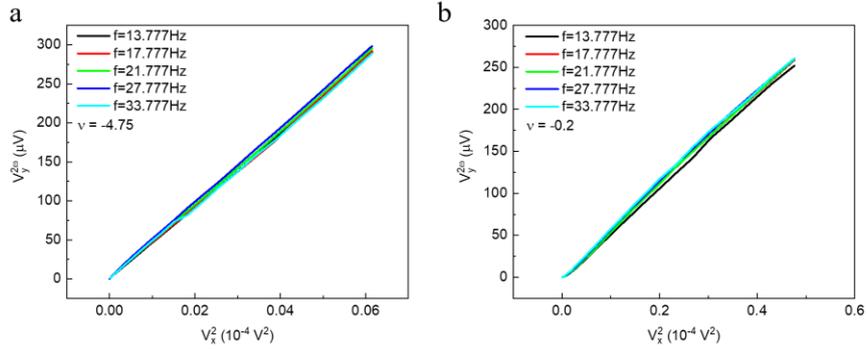

**Supplementary Figure S1. Nonlinear response under different frequencies. a,b,** Typical second-harmonic transverse voltage ($V_y^{2\omega}$) as a function of $V_x^2$ under different driving frequencies at $\nu = -4.75$ (**a**) and $\nu = -0.2$ (**b**).

2. Current dependence of $V_y^{2\omega}$

Figure S2a shows the $V_y^{2\omega}$ under different driving current at different moiré filling factor. Clearly, $V_y^{2\omega}$ depends nonlinearly on the driving current. Figure S2b shows $\frac{E_y^{2\omega}}{E_x^2}$ as a function of the moiré filling factor.

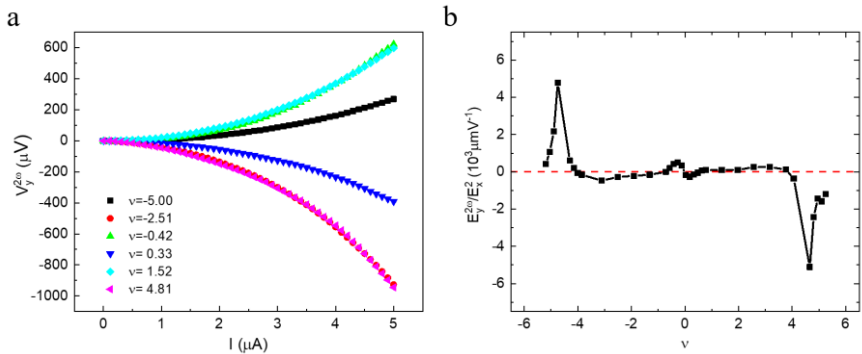

**Supplementary Figure S2. Current dependence of nonlinear response. a,** second-harmonic transverse voltage ($V_y^{2\omega}$) as a function of driving current at different moiré filling factor. **b,** $\frac{E_y^{2\omega}}{E_x^2}$

as a function of the moiré filling factor.

3. Second-harmonic longitudinal response

Since the system has $C_3$ symmetry, the longitudinal and transverse nonlinear transport should both exist and have similar order of magnitude and moiré-filling-factor dependence.

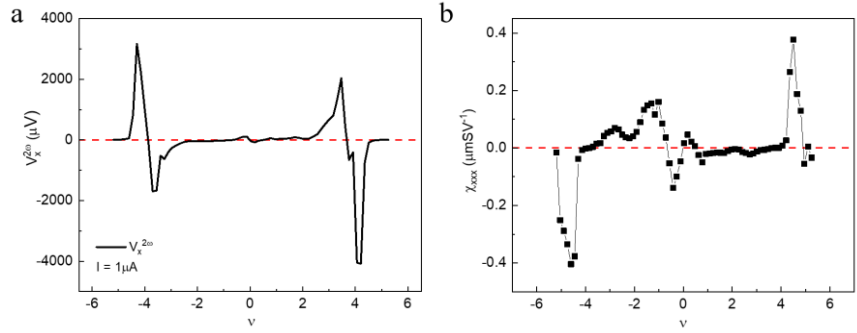

**Supplementary Figure S3. Second-harmonic longitudinal response. a,** second-harmonic longitudinal voltage ($V_x^{2\omega}$) as a function of driving current at different moiré filling factor measured in the same device as in the main text (N1). **b,** second-order longitudinal conductivity $\chi_{xxx}$ as a function of the moiré filling factor.

4. Results from another sample

We have measured nonlinear response in another TBG sample N2. Figure S4 shows the main results from N2. Figure S4a plots the moiré-filling-factor dependence of longitudinal resistance measured at 1.7 K. The peaks at moiré superlattice gap and the half filling of the moiré flat band can be clearly seen. From the position of the moiré superlattice gap, the twisted angle of this sample is estimated about 1.1 degree. Figure S4b shows moiré-filling-factor dependence of the second-harmonic response measured under 1 µA driving current. Similar to N1, the second-harmonic response peak around the moiré superlattice gap and the Dirac point. In addition, the transverse and longitudinal responses have the same moiré-filling-factor dependence, suggesting the same extrinsic origin of the two. Figure S4c and S4d shows the second-order transverse conductivity $\chi_{yxx}$ and longitudinal conductivity $\chi_{xxx}$ of N2, respectively. The magnitude of $\chi_{yxx}$ is an order smaller than $\chi_{xxx}$. This should be caused by the relative angle between the channel and the lattice. In addition, the largest magnitude of the second-harmonic response of N2 is an order smaller than that of N1. This can be explained by the larger resistivity of N2 at 1.7 K than the one of N1.

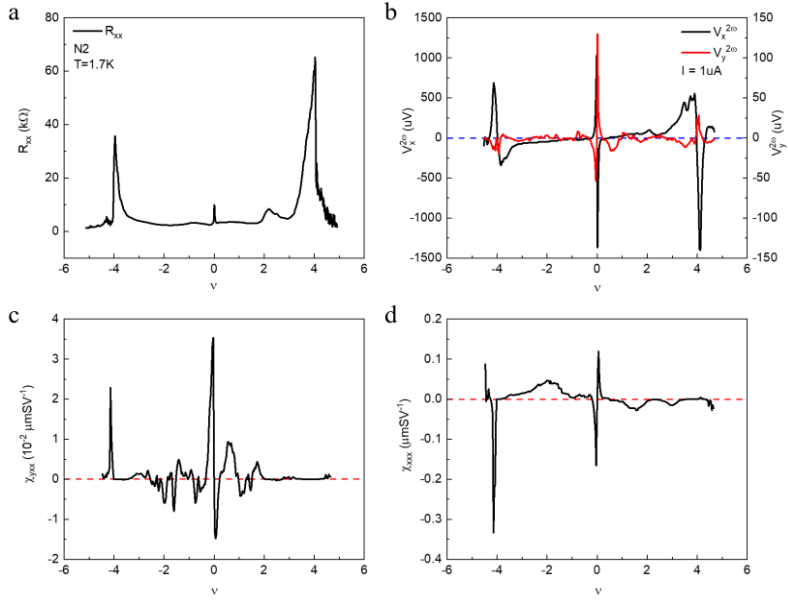

**Supplementary Figure S4. Results from another device. a,** moiré-filling-factor dependence of longitudinal resistance. **b,** second-harmonic longitudinal voltage ($V_x^{2\omega}$) and transverse voltage ($V_y^{2\omega}$) as a function of driving current at different moiré filling factor. **c,d,** second-order transverse conductivity $\chi_{yxx}$ (**c**) and longitudinal conductivity $\chi_{xxx}$ (**d**) as a function of the moiré filling factor.

5. Nonlinear conductivity

The nonlinear conductivity is defined as follows:

$$j_i^{2\omega} = \chi_{ijk} E_j E_k$$

The second-order conductivity $\chi_{ijk}$ transforms as a rand-3 tensor with index taking the value $x$ or $y$. Under the $C_3$ symmetry, we have

$$\chi_{xxx} = -\chi_{xyy} = -\chi_{yxy} = -\chi_{yyx} = \chi_1$$

$$\chi_{yyy} = -\chi_{yxx} = -\chi_{xxy} = -\chi_{xyx} = \chi_2$$

All the other components vanish. Therefore, if the electric field is at an arbitrary direction which makes $\theta$ angle with the $x$-axis, we will have

$$j_\parallel^{2\omega} = [\chi_1 \cos(3\theta) - \chi_2 \sin(3\theta)] E^2$$

$$j_\perp^{2\omega} = -[\chi_1 \sin(3\theta) + \chi_2 \cos(3\theta)] E^2$$

where $\parallel$ and $\perp$ denote the direction parallel and perpendicular to the electric field, respectively. As a result, the longitudinal nonlinear current generally exists.